\documentclass[aps,twocolumn]{revtex4-1}

\usepackage[utf8]{inputenc}
\usepackage[T1]{fontenc}
\usepackage{textcomp}
\usepackage[english]{babel}
\usepackage{graphicx}
\usepackage{latexsym,amsmath,amssymb,amsthm}
\usepackage{mathtools}
\usepackage{makeidx}
\usepackage[usenames,dvipsnames]{color}
\usepackage{natbib}
\usepackage{enumerate}

\newcommand{\PsiT}{\Psi_{\text{\tiny T}}}
\newcommand{\PsiGS}{\Psi_{\text{\tiny GS}}}
\newcommand{\Psiswf}{\Psi_{\text{\tiny SWF}}}
\newcommand{\Psifswf}{\Psi_{\text{\tiny FSWF}}}
\newcommand{\Psiaswf}{\Psi_{\text{\tiny ASWF}}}

\newcommand{\psiT}{\varphi_{\text{\tiny T}}}

\newcommand{\psiswf}{\varphi_{\text{\tiny SWF}}}
\newcommand{\psifswf}{\varphi_{\text{\tiny FSWF}}}
\newcommand{\psiaswf}{\varphi_{\text{\tiny ASWF}}}

\newcommand{\JR}{J_{\text{\tiny R}}}
\newcommand{\JS}{J_{\text{\tiny S}}}
\newcommand{\OL}{O_{\text{\tiny L}}}

\newcommand{\tpsifswf}{{\tilde{\varphi}}_{\text{\tiny FSWF}}}

\newcommand{\bpsifswf}{{\bar{\varphi}}_{\text{\tiny FSWF}}}
\newcommand{\bPsifswf}{{\bar{\Psi}}_{\text{\tiny FSWF}}}
\newcommand{\avJS}{\tilde{J}_{\text{\tiny S}}(R)}
\newcommand{\tPhi}{\tilde{\Phi}}
\newcommand{\Phipw}{{\Phi}_{\text{\tiny{pw}}}}
\newcommand{\Ntau}{N_{\tau}}

\newcommand{\GD}{\text{G}_{\text{det}}}

\newcommand{ \Hethree }{ \mbox{${}^3\text{He}$} }

\makeindex

\begin{document}
	
\title{On the Fermion Sign Problem in Imaginary-Time Projection Continuum Quantum Monte Carlo with Local Interaction}

\author{Francesco \surname{Calcavecchia}}
\email{francesco.calcavecchia@gmail.com}
\affiliation{LPMMC, UMR 5493, Bo\^ite Postale 166, 38042 Grenoble, France}
\affiliation{Institute of Physics, Johannes Gutenberg-University, Staudingerweg 7, D-55128 Mainz, Germany}
\affiliation{Graduate School of Excellence Materials Science in Mainz, Staudingerweg 9, D-55128 Mainz, Germany}

\author{Markus \surname{Holzmann}}
\email{markus.holzmann@grenoble.cnrs.fr}
\affiliation{LPMMC, UMR 5493 of CNRS, Universit{\'e} Grenoble Alpes, 38042 Grenoble, France}
\affiliation{Institut Laue Langevin, BP 156, F-38042 Grenoble Cedex 9, France}

\begin{abstract}
We use the Shadow Wave Function formalism as a convenient model to study the fermion sign problem affecting all projector Quantum Monte Carlo methods in continuum space.
We demonstrate that the efficiency of imaginary time projection algorithms decays exponentially with increasing number of particles and/or imaginary-time propagation.
Moreover, we derive an analytical expression that connects the localization of the system with the magnitude of the sign problem, illustrating this behavior through numerical results.
Finally, we discuss the computational complexity of the fermion sign problem  and methods for alleviating its severity.
\end{abstract}

\maketitle

\section*{Introduction} 
\label{sec:introduction}

The \emph{fermion sign problem} is one of the most renowned open problems in computational physics.
It consists in finding a general algorithm able to determine the exact fermionic ground state with a computational cost that grows at most polynomially with the number of simulated particles.
In fact, all  known exact algorithms for classical computers scale exponentially except for a small class of model systems \cite{PhysRevD.24.2278,Hirsch-Fey,fermionnodes,PhysRevLett.83.3116,PhysRevB.91.054413,0954-3899-37-2-025002,PhysRevB.91.241117,Entanglement,PhysRevLett.115.250601}.
In particular, Quantum Monte Carlo (QMC) methods \cite{QMCentropy2014, QMCchemrev2012, QMCrpp2011, QMCjpcm2010}, which are able to provide the exact \emph{bosonic} ground state in polynomial time for a wide range of Hamiltonians, suffer from a sign problem when applied to fermions.
In the following we will often refer to the solution of the fermion sign problem, implicitly meaning that the exact fermion ground state is obtained in polynomial time rather than in exponential time with respect to the number of particles.

In complexity theory, the class of decision problems solvable in polynomial time on a deterministic (classical) machine are called P (deterministic Polynomial time), whereas problems which can be efficiently solved by probabilistic algorithms are called BPP (Bounded-error Probabilistic Polynomial time).
In respect to this terminology, general fermionic simulations suffering from the sign problem seem to remain outside the P/BPP classes.
This is a reminiscence of  Non-deterministic Polynomial Complete (NPC) problems, a set of hundreds of apparently different problems that, despite many efforts, have not been solved yet.
The most famous of such problems is the Travelling Salesman problem, formulated in 1930.
Notably, it has been found that all these problems are mappable one into the other, so that the solution of one of them would imply the solution of all of them \cite{book:algorithms}.
The fact that it has not been possible to solve not even one of them, justifies the common belief that these problem are intractable.
As a matter of fact, NPC problems are so firmly believed to be intractable, that all classical encryption schemes rely on this conjecture.
However, despite the importance of this conjecture (known in the literature as $\text{NP} \neq \text{P}$ hypothesis), a proof is still missing.

\begin{figure}[h]
	\centering
	\includegraphics[width=8.5cm]{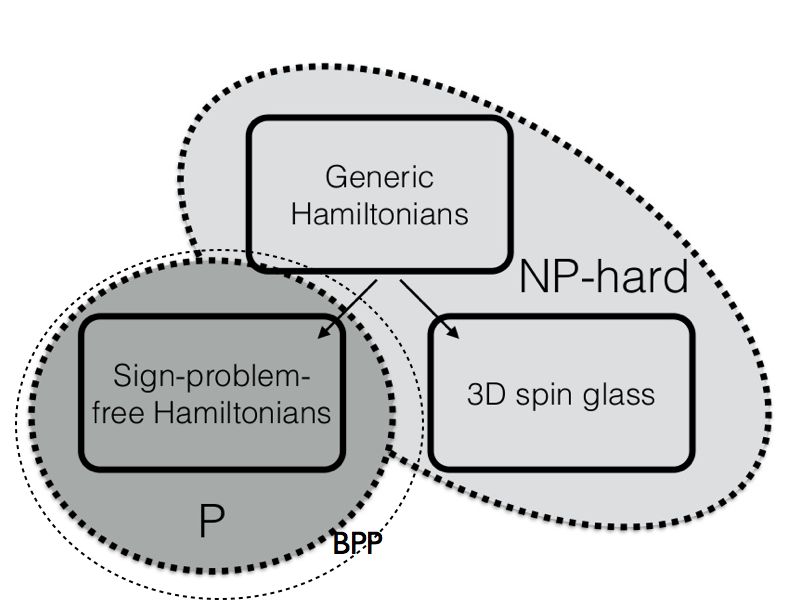} 
	\caption{Schematic representation of computational complexity of QMC simulations.}
	\label{fig:P_vs_NP-hard_Hamiltonian}
\end{figure}

In 2005, Troyer and Wiese provided a demonstration of the NP-hardness of the Monte Carlo (MC) sign problem \cite{PhysRevLett.94.170201} for a specific Hamiltonian.
The NP-hard problems are a class of problems which are mappable in polynomial time into NPC so that solving any NP-hard problem would provide the solution to all  NPC problems.
In this sense NP-hard problems are said to be the hardest ones, as they are at least as hard as any other NPC problem (see Fig~\ref{fig:P_vs_NP-hard_Hamiltonian}).

In the present paper, we  focus on fermionic QMC simulations of continuum systems with local interactions, e.g. non-relativistic electrons interacting via static Coulomb potential, and discuss in detail the corresponding fermion sign problem.
In particular, we study the performance of the imaginary-time projection techniques for these systems, which provide polynomial scaling solutions to the corresponding bosonic problems. We give a general proof of the exponential scaling of the efficiency of such algorithms, both in fermion number and projection time. Further, we explicitly show that
localized orbitals can drastically reduce the sign problem.
Our discussion is based on the Shadow Wave Function (SWF) formalism \cite{PhysRevLett.60.1970}.
If the fermion sign problem can be solved for the SWF, such a solution will be extendable to all the other QMC methods.
Whereas, if SWF can be proved to fall in the NP-hard class,  the same will apply to the other imaginary-time projections QMC methods.

We remark that, even if a problem is NP-hard, one can obtain approximate solutions, and work on the improvement of the approximations.
For example, even though the Travelling Salesman problem cannot be solved exactly for many problems of practical interest, there are several  methods that provide excellent approximations \cite{LAPORTE1992231,doi:10.1287/opre.28.3.694}.
Concerning the fermion sign problem, remarkable progress has been obtained over the recent years  by employing more flexible trial wave functions and improved  optimization procedures, which systematically reduce the deviations from the exact ground state \cite{PhysRevLett.98.110201,He3,PhysRevB.91.115106,Luechow:_CI_in_FNDMC}.

A different strategy which we follow in this paper is to alleviate the sign problem, where one aimes to reduce the pre-factor of the ultimate exponential decay in the signal of the desired quantities, similar to released-node algorithm \cite{ReleaseNode} or different exact fermion simulations \cite{prospects_release_node_QMC,PhysRevE.50.3220,PhysRevLett.85.3547,PhysRevB.71.155115,PhysRevLett.91.186402,Saverio1,Saverio2}.
In order to compare different approaches in the following, we refer to the Monte Carlo \emph{efficiency}, defined as
\begin{equation}
	\eta \equiv \frac{1}{\text{cpu time} \times \text{var}} \, ,
\end{equation}
where $\text{var}$ is the variance of the computed quantity.
The SWF is a perfect tool-bench for a systematic study of the efficiency of different strategies, as the severity of the sign problem can be controlled by kernel parameters, and the method itself is considerably cheaper than most of imaginary-time projection methods.

This paper is structured as follows.
In Sec. \ref{sec:shadow_wave_function} we introduce the Shadow Wave Function formalism, and then explain its connection with Diffusion Monte Carlo and Path Integral Ground State in Sec. \ref{sec:connection}.
We will characterize its fermion sign problem when using a Slater determinant of simple plane waves in Sec. \ref{sec:sign_problem}, and generalize this result to orbitals of any kind in Sec. \ref{sec:generalization_to_any_kind_of_slater_determinant}.
We will then review some general method that have been proposed to tackle the fermion sign problem in Sec. \ref{sec:alleviation}, and finally discuss about its computational complexity in Sec. \ref{sec:complexity}.
The conclusion are drawn in Sec. \ref{sec:conclusion}.

\section{The Shadow Wave Function Formulation} 
\label{sec:shadow_wave_function}
The Shadow Wave Function (SWF) is a class of variational trial wave functions which, by embedding an integral, can profit from a great flexibility \cite{PhysRevLett.60.1970,PhysRevB.38.4516}. 
In general, the SWF can be written as
\begin{equation}
\begin{split}
	\Psiswf(R) &= \JR(R) \int dS \, \Xi(R,S) \psiT(S) \\
			   &= \int dS \psiswf(R,S) \label{eq:generic_SWF}
\end{split}
\end{equation}
where $R \equiv (\mathbf{r}_1, \mathbf{r}_2, \dots \mathbf{r}_N)$ represents the particle coordinates, $S$ labels some $3N$-dimensional auxiliary coordinates, $\JR$ is a Jastrow, $\psiT$ is a chosen trial wave function, and
\begin{equation}
	\Xi(R,S) = e^{-C(R-S)^2}
\end{equation}
is the so-called \emph{kernel}.
The SWF efficiently removes large part of the bias introduced by the underlying trial wave function, in particular close
to phase transitions or in inhomogeneous systems  \cite{PhysRevLett.72.2589, PhysRevB.56.5909, PhysRevB.69.024203, PhysRevLett.102.255302, KalosReatto1995, calcavecchia:swf_H2}

The application of the SWF to fermions requires the fulfilment of the Fermi-Dirac statistics
by the use of an antisymmetric form for spin-like particles.
In the following we
 split the trial wave function into the product of a symmetrical part, the Jastrow factor, times an antisymmetrical part, 
 usually taken as a Slater determinant.

A straightforward antisymmetrization of  Eq. \eqref{eq:generic_SWF} gives the Fermionic Shadow Wave Function (FSWF) form
\begin{equation}
\begin{split}
	\Psifswf(R) &= \JR(R) \int dS \, \Xi(R,S) \JS(S) \Phi(S) \\
	            &= \int dS \psifswf(R,S)
\end{split}
\end{equation}
where $\Phi$ represents a Slater determinant and $\JS$ a Jastrow. 
Unfortunately, the FSWF introduces a sign problem \cite{PhysRevE.90.053304} which emerges by the fact that the product \mbox{$\psifswf^*(R,S_1) \times \psifswf(R,S_2)$} is not necessarily positive.

The sign problem of FSWF is avoided using the Antisymmetric Shadow Wave Function (ASWF)
\begin{equation}
\begin{split}
	\Psiaswf(R) &= \JR(R) \Phi(R) \int dS \, \Xi(R,S) \JS(S) \\
	            &= \int dS \psiaswf(R,S) \, ,
\end{split}
\end{equation}
since \mbox{$\psiaswf^*(R,S_1) \times \psiaswf(R,S_2)$} is garantied to be positive for any $R$, $S_1$ and $S_2$.

Typically, FSWF is considered of higher quality and expected to yield lower variational energies than ASWF.

\section{Connection Between SWF, DMC, and PIGS} 
\label{sec:connection}
The SWF can be regarded as a prototype method for any kind of imaginary-time projection method, since the kernel has the form of an approximated Green's function with $C$ proportional to the inverse imaginary-time propagation $\tau$.

Let us consider a trial wave-function $\psiT$ which is not orthogonal to the exact ground state.
It is well-known that a propagation in imaginary time will eventually project it to the exact ground state, i.e.
\begin{equation}
   e^{-\tau H} \mid \PsiT \rangle \xrightarrow{\tau \rightarrow \infty} \mid \PsiGS \rangle \label{eq:imaginary-time_propagation}
\end{equation}
In order to build a concrete algorithm built upon such property, we make use of the Suzuki-Trotter formula and write
\begin{equation}
   e^{-\tau (T+V)} \simeq e^{-\tau T} \, e^{-\tau V} \, , \label{eq:suzuki-trotter}
\end{equation}
which is exact in the limit $\tau \rightarrow 0$.

In order to conciliate the necessity of having a large $\tau$ (Eq. \eqref{eq:imaginary-time_propagation}) with the Suzuki-Trotter approximation which requires a small $\tau$, we can break the propagation into several small ones.
Given a $\Ntau>1$ such that that $\delta \tau \equiv \tau/\Ntau$ is small enough for the approximation in Eq. \eqref{eq:suzuki-trotter} to hold, the operators $e^{-\delta \tau T} e^{-\delta \tau V}$ can be applied iteratively to the trial wave function to project out the ground state wave function 
\begin{equation}
\begin{split}
  \mid \PsiGS \rangle \simeq   \mid \PsiT[\tau] \rangle & \equiv  e^{-\tau H} \mid \PsiT[0] \rangle \\
  &\simeq \left( \prod_{i=1}^{\Ntau} e^{-\delta \tau T} e^{-\delta \tau V} \right) \mid \PsiT[0] \rangle \\
\end{split}
\end{equation}
The SWF has the functional form of  single propagation $\delta \tau$, since the kinetic energy propagator is a diffusor term in the coordinate space, i.e.
\begin{equation}
   \langle R \mid e^{-\delta \tau T} \mid R^{\prime} \rangle \propto e^{-\frac{(R^{\prime}-R)^2}{4 \delta \tau}} 
\end{equation}
Using a variation of the Suzuki-Trotter approximation which splits the potential propagator on the left and on the right symmetrically
\begin{equation}
   e^{-\delta \tau (T+V)} \simeq e^{-\frac{\delta \tau}{2} V} \, e^{-\tau T} \, e^{-\frac{\delta \tau}{2} V} \, ,
\end{equation}
we can write (assuming that the potential $V$ is real and local)
\begin{equation}
\begin{split}
   \psiT[\delta \tau](R) &= \langle \PsiT[\delta \tau] \mid R \rangle \\
            &= \langle \PsiT[0] \mid e^{-\frac{\delta \tau}{2} V} \, e^{-\delta \tau T} \, e^{-\frac{\delta \tau}{2} V} \mid R \rangle \\
            &= \int dS \, \langle \PsiT[0] \mid S \rangle \langle S \mid e^{-\frac{\delta \tau}{2} V} \, e^{-\delta \tau T} \, e^{-\frac{\delta \tau}{2} V} \mid R \rangle  \\
            &= \int dS \, \psiT(S) \, e^{-\frac{\delta \tau}{2} V(S)} \, e^{-\frac{(R-S)^2}{4 \delta \tau}} \, e^{-\frac{\delta \tau}{2} V(R)}
\end{split}
\end{equation}
The generic form of the SWF (Eq. \eqref{eq:generic_SWF}) is the one obtained by mapping
\begin{equation}
   \left\{
   \begin{array}{l}
      \frac{1}{4 \delta \tau} \mapsto C \\
      \psiT(S) \, e^{-\frac{\delta \tau}{2} V(S)} \mapsto \psiT(S) \\
      e^{-\frac{\delta \tau}{2} V(R)} \mapsto \JR(R)
   \end{array}
   \right.
\end{equation}

In contrast to projector Monte Carlo methods, e.g. Diffusion Monte Carlo (DMC) or Path Integral Ground State Monte Carlo (PIGS),
SWF can be considered as a single step of a chain of small imaginary-time propagations.
As a consequence,  SWF is in general not exact, even though it often captures most of the corrections of the imaginary-time propagation, while retaining a low-computational cost.
Further, SWF remains a explicit trial wave function subject to the Rayleigh-Ritz variational principle, and $\delta \tau$ (corresponding to  $C$) can be regarded as a variational parameter at variance to
 projector Monte Carlo methods which must be extrapolated to the limit $\delta \tau \to 0$.

\section{The Sign Problem of the Fermionic Shadow Wave Function} 
\label{sec:sign_problem}
The sign problem of the Fermionic Shadow Wave Function has been explored numerically  in Ref.~\cite{PhysRevE.90.053304}.
In this section we are going to introduce two different approaches which justify qualitatively and quantitatively its occurrence.

We will assume that the orbitals of the Slater determinant are simple plane waves throughout this whole section and generalize the results to any kind of orbitals in Sec. \ref{sec:generalization_to_any_kind_of_slater_determinant}.

\subsection{Ratio with a positive-definite distribution} 
\label{sub:ratio_with_a_positive_definite_distribution}
The expectation value of an operator $O$ computed averaging over a Shadow Wave Function writes
\begin{equation}
\label{eq:Oswf}
\begin{split}
	\langle O \rangle &= \frac{\int dR \, dS_1 \, dS_2 \, \psiswf^*(R,S_1) \, O \, \psiswf(R,S_2)}{\int dR \, dS_1 \, dS_2 \, \psiswf^*(R,S_1) \psiswf(R,S_2)} \\
	&= \frac{\int dR \, dS_1 \, dS_2 \, \psiswf^*(R,S_1) \psiswf(R,S_2) \, \OL(R,S_2)}{\int dR \, dS_1 \, dS_2 \, \psiswf^*(R,S_1) \psiswf(R,S_2)}
\end{split}
\end{equation}
where
\begin{equation}
	\OL(R,S) \equiv \frac{O \psiswf(R,S)}{\psiswf(R,S)} \, .
\end{equation}

If we denote by $\rho(R,S_1,S_2)$ the probability density function (pdf) that we intend to sample from, and introduce the corresponding weight
\begin{equation}
	w(R,S_1,S_2) = \frac{\psiswf^*(R,S_1) \psiswf(R,S_2)}{\rho(R,S_1,S_2)} \, ,
\end{equation}
then we can recast Eq. \eqref{eq:Oswf} as
\begin{equation}
	\langle O \rangle = \frac{\int dR \, dS_1 \, dS_2 \, \rho(R,S_1,S_2) \, w(R,S_1,S_2) \, \OL(R,S_2)}{\int dR \, dS_1 \, dS_2 \, \rho(R,S_1,S_2) \, w(R,S_1,S_2)} \, .
\end{equation}

If the product $\psiswf^*(R,S_1) \times \psiswf(R,S_2)$ is positive-definite, we can chose $\rho$ such that $w=1$ and no sign problem will occur.

However, if this is not the case, a typical choice is
\begin{equation}
	\rho(R,S_1,S_2) \equiv \left| \psiswf^*(R,S_1) \psiswf(R,S_2) \right|
\end{equation}
and therefore
\begin{equation}
	w(R,S_1,S_2) = \text{sign}(\psiswf^*(R,S_1) \psiswf(R,S_2)) = \pm 1 \, .
\end{equation}
introducing a sign problem.
The expectation value of $O$ is then
\begin{equation}
	\langle O \rangle = \frac{\langle w \, \OL \rangle_{\rho}}{\langle w \rangle_{\rho}}
\end{equation}
where by $\langle \dots \rangle_{\rho}$ we mean the average resulting from sampling the pdf $\rho$.

Let us now focus on $\langle w \rangle_{\rho}$.

In the case of PIGS with a projection time $\tau$ or Path-Integral Monte Carlo at finite temperature $T=1/\tau$, $\langle w \rangle_{\rho}$ is equal to the ratio between the fermionic and the bosonic partition functions \cite{CeperleyPathIntergralForFermionsReview1996} (where the bosonic system is defined by the positive-definite weight $|w|$)
and therefore
\begin{equation}
	\label{eq:w_PI}
	\langle w \rangle_{\rho} = e^{-\tau N \Delta F}
\end{equation}
where $N$ is the number of particles and $\Delta F \ge 0$ is the free energy difference per particle between the fermionic and the bosonic system.
Relation \eqref{eq:w_PI} explains the exponential decay in efficiency, as 
\begin{equation}
	\label{eq:var(w)}
	\frac{\sigma(\langle w \rangle_{\rho})}{\langle w \rangle_{\rho}} = \sqrt{\frac{\langle w^2 \rangle_{\rho} - \langle w \rangle^2_{\rho}}{M \langle w \rangle_{\rho}^2}} = \sqrt{\frac{1/\langle w \rangle^2_{\rho}-1}{M}} \sim \frac{e^{\tau N \Delta F}}{\sqrt{M}}
\end{equation}
where $M$ is the number of sampled points.
Since the relative error of $\langle O \rangle$ is given by the sum of the relative errors of $\langle w \OL \rangle_{\rho}$ and $\langle w \rangle_{\rho}$, we can see that Eq. \eqref{eq:var(w)} is sufficient to explain the exponential decay of the efficiency of any observable with $N$ and $\tau$.

In the case of  SWF, we can evaluate $\langle w \rangle_{\rho}$ explicitly, by making two assumptions:
\begin{enumerate}[i]
	\item The bosonic system is represented by an ASWF;
	\item Correlation factors (Jastrow) do not play a crucial role and therefore can be omitted.
\end{enumerate}
Under these assumptions, $\langle w \rangle_{\rho}$ associated to $\Psifswf$
\begin{equation}
	\langle w \rangle_{\text{FSWF}} = \frac{\int dR \, dS_1 \, dS_2 \, \psifswf^*(R,S_1) \, \psifswf^*(R,S_2) }{\int dR \, dS_1 \, dS_2 \, \psiaswf^*(R,S_1) \, \psiaswf^*(R,S_2)}
\end{equation}
can be simplified using
\begin{equation}
\begin{split}
	\psifswf(R,S) &\simeq e^{-C (R-S)^2} \det(e^{i \mathbf{k}_{\alpha} \cdot \mathbf{s}_{\beta} }) \\
	\psiaswf(R,S) &\simeq e^{-C (R-S)^2} \det(e^{i \mathbf{k}_{\alpha} \cdot \mathbf{r}_{\beta} }) \, .
\end{split}
\end{equation}
where exactly $N$ wave vectors $\mathbf{k}$ are occupied in the Slater determinant, and where we assume a spin-polarized system with $N$ fermions for simplification.
We then have
\begin{equation}
\begin{split}
	\int dS \, & e^{-C (R-S)^2} \det(e^{i \mathbf{k}_{\alpha} \cdot \mathbf{s}_{\beta} }) = \\
	& \left( \frac{\pi}{C} \right)^{\frac{3N}{2}} e^{-\frac{\sum_{i=1}^N \mathbf{k}_i^2}{4C}} \det(e^{i \mathbf{k}_{\alpha} \cdot \mathbf{r}_{\beta} })
\end{split}
\end{equation}
so that we can integrate out $dS_1 \, dS_2$, and obtain
\begin{equation}
\begin{split}
	\langle w \rangle_{\text{FSWF}} &\simeq \frac{ e^{-\frac{\sum_{i=1}^N \mathbf{k}_i^2}{4C}} \int dR \, \det(e^{-i \mathbf{k}_{\alpha} \cdot \mathbf{r}_{\beta} }) \, \det(e^{i \mathbf{k}_{\alpha} \cdot \mathbf{r}_{\beta} }) }{ \int dR \, \det(e^{-i \mathbf{k}_{\alpha} \cdot \mathbf{r}_{\beta} }) \, \det(e^{i \mathbf{k}_{\alpha} \cdot \mathbf{r}_{\beta} }) } \\
	&= e^{-\frac{\sum_{i=1}^N \mathbf{k}_i^2}{4C}} \, .
\end{split}
\end{equation}
In the thermodynamic limit we then get
\begin{equation}
	\langle w \rangle_{\text{FSWF}} \propto e^{- \frac{N}{C} \rho^{2/3}} \, .
\end{equation}
where $\rho$ is the density. 
Therefore, the efficiency of a QMC simulation for computing $\langle O \rangle$ employing the FSWF writes
\begin{equation}
	\eta \propto e^{- \frac{N}{C} \rho^{2/3}} \, .
\end{equation}
Assumptions i and ii may be relaxed for situations where reweighting is possible, but in 
subsection \ref{sub:difference_with_a_positive_definite_distribution} we will derive the same scaling without 
relying on them at all.

\subsection{Difference with a positive-definite distribution} 
\label{sub:difference_with_a_positive_definite_distribution}

In the spirit of the \emph{control variates} technique \cite{doi:10.1287/opre.1.5.263, Kalos:MC}, we recast the FSWF as
\begin{equation}
\begin{split}
	\Psifswf(R) =& \int dS \, \left( \psifswf(R,S) - \tpsifswf(R,S) \right) \\
	             & + \int dS \, \tpsifswf(R,S) \, , \label{eq:control_variates}
\end{split}
\end{equation}
with
\begin{equation}
		\tpsifswf(R,S) = \JR(R) \, \avJS \, \Xi(R,S) \Phi(S)
\end{equation}
We can now chose the local normalization factor, $\avJS$, such  that
\begin{equation}
		\int dS \, \tpsifswf(R,S) = \int dS \, \psifswf(R,S) 
\end{equation}
We see that $\tpsifswf$ is a fermionic shadow wave function in which the shadow-shadow correlation has been replaced by an effective local Jastrow $\avJS$. Although needed for our proof,
the reader should bear in mind that the normalization factor involved cannot be easily estimated, its computation itself will lead to a sign problem.

We now integrate out the shadows, in order to eliminate the sign problem in the second integral of Eq. \eqref{eq:control_variates}:
\begin{equation}
	\begin{split}
		\bPsifswf(R) \equiv& \, \int dS \, \tpsifswf(R,S) \\
		                    =& \, \JR(R) \, \avJS \, \Phi(R) \, \left( \frac{\pi}{C} \right)^{\frac{3N}{2}} \,  e^{-\frac{\sum_{i=1}^N \mathbf{k}_i^2}{4C}}\\ 
		=& \, \JR(R) \, \avJS \, \Phi(R) \, e^{-\frac{\sum_{i=1}^N \mathbf{k}_i^2}{4C}} \, \int dS \, \Xi(R,S) \\
		             =& \, \int dS \, \bpsifswf(R,S)
	\end{split}
\end{equation}
where we have formally reintroduced the shadows.
We can see that $\bPsifswf$ can be regarded as the closest ASWF to the given FSWF, i.e. a "bosonized" FSWF.
Notice that $\bPsifswf(R)$ does not contain a Slater determinant evaluated on the shadow coordinates anymore, hence this wave function is not affected by the sign problem.

The FSWF can now be written as
\begin{equation}
\begin{split}
	\Psifswf(R) =& \int dS \, \left[ \psifswf(R,S) - \tpsifswf(R,S) \right. \\
	             & \hspace{3.05cm} \left. + \bpsifswf(R,S) \right] \\
				=& \, \JR(R) \, \int dS \, \Xi(R,S) \, \avJS \times \\
				 & \left[ \Phi(S) \, \left( \frac{\JS(S)}{\avJS} - 1 \right) + \Phi(R) \, e^{-\frac{\sum_{i=1}^N \mathbf{k}_i^2}{4C}} \right] \label{eq:FSWF_after_control_variates}
\end{split}
\end{equation}
If the term $\left( \frac{\JS(S)}{\avJS} - 1 \right)$ were  zero, we would have solved the sign problem. 

Let us now assume that we are able to sample from the signal, i.e.
\begin{equation}
	\rho(R,S_1,S_2) = \bpsifswf^*(R,S_1) \, \bpsifswf(R,S_2) \, ,
\end{equation}
then the weight corresponding to our original sampling, Eq.~(\ref{eq:FSWF_after_control_variates}), writes 
\begin{equation}
\begin{split}
	w(R,S_1,S_2) =& \left[ 1 + \frac{\Phi(S_1)}{\Phi(R)} \, \left( \frac{\JS(S_1)}{\avJS} - 1 \right) \, e^{\frac{\sum_{i=1}^N \mathbf{k}_i^2}{4C}} \right ] \\
	             & \times \left[ 1 + \frac{\Phi(S_2)}{\Phi(R)} \, \left( \frac{\JS(S_2)}{\avJS} - 1 \right) \, e^{\frac{\sum_{i=1}^N \mathbf{k}_i^2}{4C}} \right ]
\end{split}
\end{equation}
which can be seen as
\begin{equation}
	\text{weight} = \text{signal} + \text{noise}
\end{equation}
where signal$=1$.
Hence, we require that
\begin{equation}
	\left| \langle \text{noise} \rangle \right| < \varepsilon \left| \langle \text{signal} \rangle \right|
\end{equation}
where $\varepsilon<1$ determines the final error of the calculation.
We get
\begin{equation}
	\begin{split}
		\left| \left\langle \frac{\Phi(S_1)}{\Phi(R)} \, \left( \frac{\JS(S_1)}{\avJS} - 1 \right) \right\rangle \, e^{\frac{\sum_{i=1}^N \mathbf{k}_i^2}{4C}} \right. & \, \\
		+ \left\langle \frac{\Phi(S_2)}{\Phi(R)} \, \left( \frac{\JS(S_2)}{\avJS} - 1 \right) \right\rangle \, e^{\frac{\sum_{i=1}^N \mathbf{k}_i^2}{4C}} & \, \\
		+ \left\langle \frac{\Phi(S_1)\Phi(S_2)}{\Phi^2(R)} \, \left( \frac{\JS(S_1)}{\avJS} - 1 \right) \right. \\
	   \left. \left. \times \left( \frac{\JS(S_2)}{\avJS} - 1 \right) \right\rangle \, e^{2 \frac{\sum_{i=1}^N \mathbf{k}_i^2}{4C}} \right| & < \varepsilon \label{eq:requirement_no_sign_problem}
	\end{split}
\end{equation}

Equation \eqref{eq:requirement_no_sign_problem} is a necessary condition to avoid the sign problem, and it contains all the informations that we are looking for.

In particular, we would like to know how the sign problem scales with the number of particles.
For that we use the following reasoning: Suppose that we have performed a calculation with $N_0$ particles which provided us 
an estimate for our observable within a given  error bar and the corresponding efficiency $\eta_0$.
We then change the number of particles to $N=\kappa N_0$.
To obtain the same accuracy, we must require that the noise term in the weight is the same, i.e. $\langle \text{noise} \rangle = \langle \text{noise}_0 \rangle$.
From Eq. \eqref{eq:requirement_no_sign_problem}, we can read a dependence from the number of particles in the terms $e^{-\frac{\sum_{i=1}^N \mathbf{k}_i^2}{4C}} \simeq e^{-\frac{N}{C} \rho^{2/3}}$.
Therefore, the averages in Eq. \eqref{eq:requirement_no_sign_problem} must be decreased by a factor $e^{-\kappa \frac{N_0}{C} \rho^{2/3}}$.
Assuming the variance of the noise integrand to be independent of $N$, we have
		$\sigma \propto  (\text{number of sampled points})^{-1/2} 
		       \propto  (\text{cpu time})^{-1/2}$
and we can conclude that
	$\log \eta/\eta_0 \propto - \kappa$.
Similarly, we can derive the dependencies for $C$ and $\rho$, which gives
\begin{equation}
	\log \eta/\eta_0 \propto- \frac{N}{C} \rho^{2/3}  \, .
\end{equation}
yielding the result of the previous subsection under more general assumptions.

\begin{figure}
	\centering
		\includegraphics[width=8.5cm]{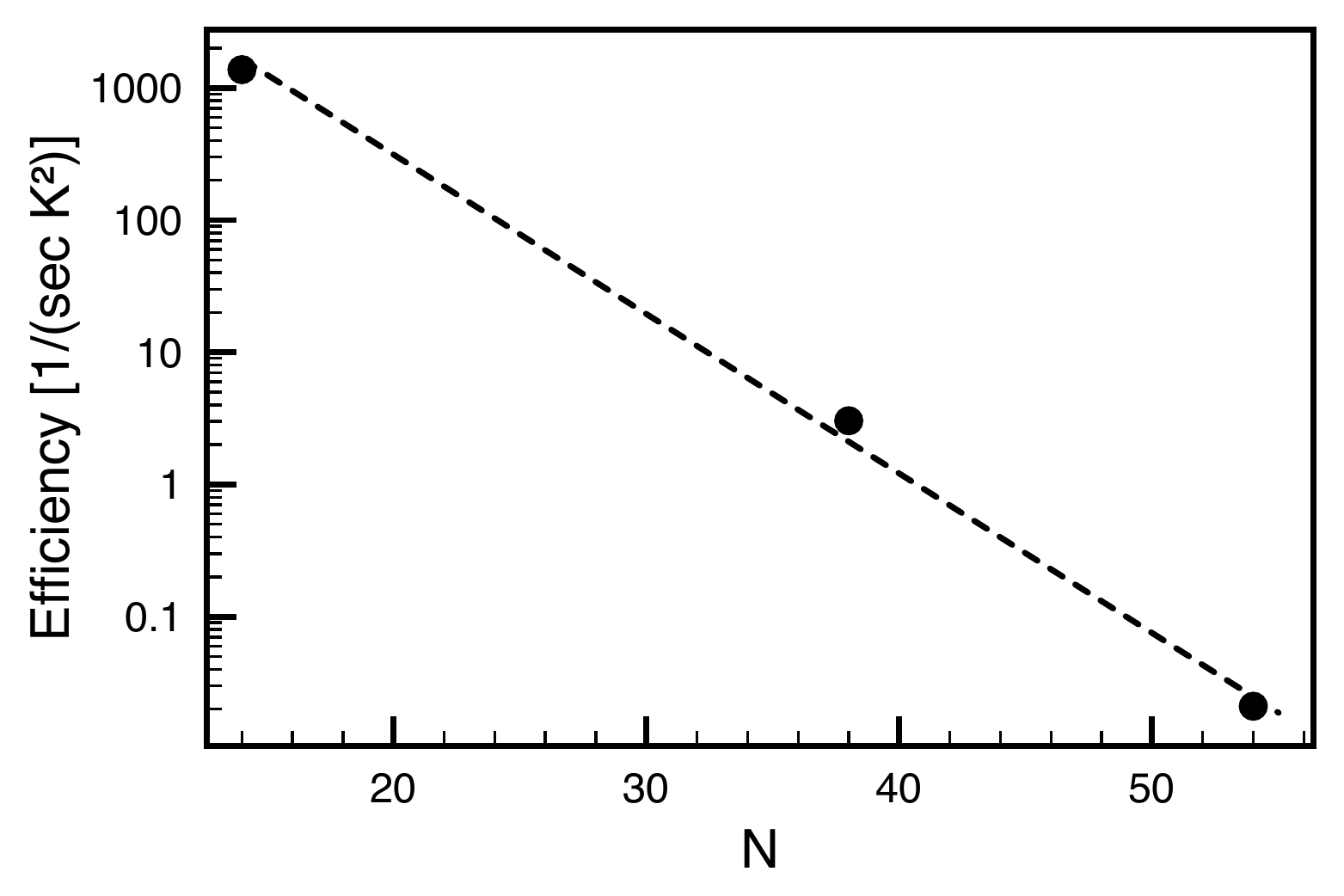} 
	\caption{Efficiency of a VMC simulation of liquid \Hethree that employs the FSWF \cite{PhysRevE.90.053304}, fitted with an exponential function $\sim e^{-kN}$, where $k$ is a constant.}
	\label{fig:fit-N}
\end{figure}

\begin{figure}
	\centering
		\includegraphics[width=8.5cm]{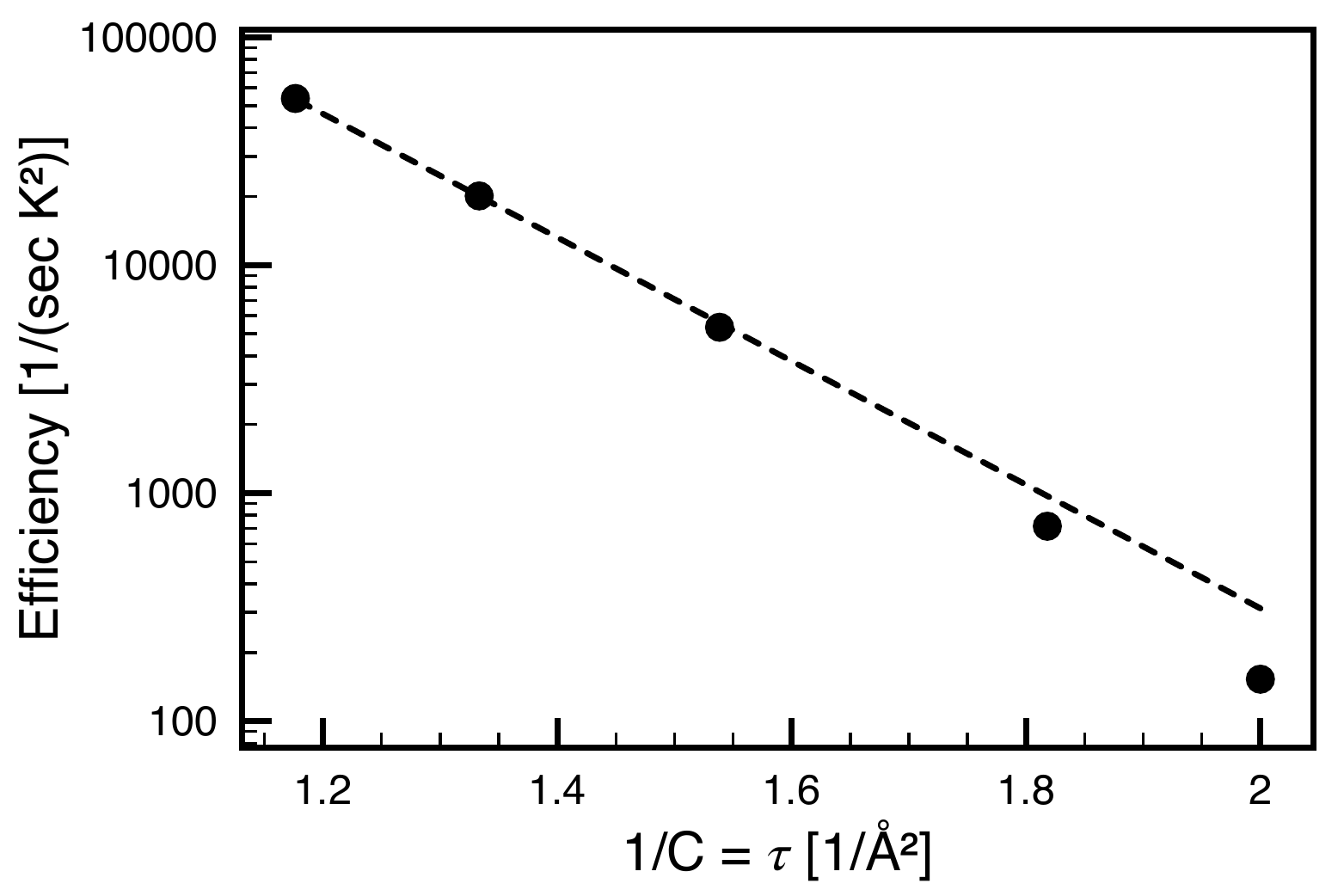} 
	\caption{Efficiency of a VMC simulation of liquid \Hethree that employs the FSWF \cite{PhysRevE.90.053304}, fitted with an exponential function $\sim e^{-k/C}$, where $k$ is a constant.}
	\label{fig:fit-C}
\end{figure}

These prediction are confirmed by the numerical results reported in \cite{PhysRevE.90.053304}
using FSWF trial wave functions to compute the VMC energy of  unpolarized liquid \Hethree ($\rho=0.016588~\text{\AA}^{-3}$) in three dimensions.
In Figures \ref{fig:fit-N} and \ref{fig:fit-C} these datas are fitted with an exponential function, demonstrating the exponential dependency on $N$ and $C$.

\section{Generalization to Any Kind of Orbitals} 
\label{sec:generalization_to_any_kind_of_slater_determinant}
In this section we are going to generalize the dependence of the efficiency on $N$, $C$, and $\rho$ to the more general case of Slater determinants which use any kind of orbitals.

To accomplish this result it is sufficient to work in Fourier space.
The matrix elements of the Slated determinant can be expressed as an integral  of a product of matrices over the N-particle momentum space:
\begin{equation}
	\begin{split}
		\phi_{\alpha}(\mathbf{r}_{\beta}) =& \int d\mathbf{k}_{\alpha} \, e^{-i \mathbf{k}_{\alpha} \cdot \mathbf{r}_{\beta}} \, \tilde{\phi}_{\alpha}(\mathbf{k}_{\alpha}) \\
		\sim & \int dK \, \sum_{\gamma=1}^{N} \, \left( e^{-i \mathbf{k}_{\gamma} \cdot \mathbf{r}_{\beta}} \right) \, \left( \tilde{\phi}_{\alpha}(\mathbf{k}_{\gamma}) \, \delta_{\gamma \alpha} \right) \\
		=& \int dK \, ( e^{-i \mathbf{k}_{\gamma} \cdot \mathbf{r}_{\beta}} ) \cdot ( I_{\tilde{\phi}_{\alpha}(\mathbf{k}_{\gamma})} ) \, ,
	\end{split}
\end{equation}
where $\tilde{\phi}(\mathbf{k})$ is the Fourier transform of the orbital $\phi(\mathbf{r})$, and
$$ I_{\tilde{\phi}_{\alpha}(\mathbf{k}_{\gamma})} \equiv \tilde{\phi}_{\alpha}(\mathbf{k}_{\gamma}) \, \delta_{\gamma \alpha} \; .$$
Therefore
\begin{equation}
\begin{split}
	\det(\phi_{\alpha}(\mathbf{r}_{\beta})) \sim& \int dK \, \det( e^{-i \mathbf{k}_{\gamma} \cdot \mathbf{r}_{\beta}} ) \, \det( I_{\tilde{\phi}_{\alpha}(\mathbf{k}_{\gamma})} ) \\
	=& \int dK \, \det( e^{-i \mathbf{k}_{\gamma} \cdot \mathbf{r}_{\beta}} ) \, \prod_{\gamma=1}^N \tilde{\phi}_{\gamma}(\mathbf{k}_{\gamma})
\end{split}
\end{equation}
In the following we will use the notation
\begin{equation}
	\begin{split}
		\Phi(R) & \equiv \det({\phi}_{\alpha}(\mathbf{r}_{\beta})) \\
		\tPhi(K) & \equiv \prod_{\gamma=1}^N \tilde{\phi}_{\gamma}(\mathbf{k}_{\gamma})
	\end{split}
\end{equation}
and
\begin{equation}
	\begin{split}
		\Phipw(R,K) \equiv \det(e^{-i \mathbf{k}_{\alpha} \cdot \mathbf{r}_{\beta}})
	\end{split}
\end{equation}

Following the idea used in subsection \ref{sub:difference_with_a_positive_definite_distribution}, we write:
\begin{equation}
	\begin{split}
		\psifswf(R,S,K) = & \JR(R) \, \Xi(R,S) \, \JS(S) \, \tPhi(K) \, \Phipw(R,S) \\
		\tpsifswf(R,S,K) = & \JR(R) \, \avJS \, \Xi(R,S) \, \tPhi(K) \, \Phipw(R,S) \\
		\bpsifswf(R,S,K) = & \JR(R) \, \avJS \, \Xi(R,S) \, e^{-\frac{\sum_{i=1}^N \mathbf{k}_i^2}{4C}}  \\
		              & \hspace{1cm} \times \tPhi(K) \, \Phipw(R,K) \, .
	\end{split}
\end{equation}

Eq. \eqref{eq:FSWF_after_control_variates} can be recasted as
\begin{equation}
\begin{split}
	\Psifswf(R) \sim & \, \JR(R) \, \int dS \, dK \, \Xi(R,S) \, \avJS \times \\
				 & \left[ \tPhi(K) \, \Phipw(S) \, \left( \frac{\JS(S)}{\avJS} - 1 \right)  \right. \\
				 & \left. \qquad + \tPhi(K) \, \Phipw(R) \, e^{-\frac{\sum_{i=1}^N \mathbf{k}_i^2}{4C}} \right] \; ,
\end{split}
\end{equation}
while Eq. \eqref{eq:requirement_no_sign_problem} writes
\begin{equation}
	\begin{split}
		\left| \left\langle \frac{\Phipw(S_1,K_1)}{\Phipw(R,K_1)} \, \left( \frac{\JS(S_1)}{\avJS} - 1 \right) \, e^{\frac{\sum_{i=1}^N \mathbf{k}_i^2}{4C}} \right\rangle \right. & \, \\
		+ \left\langle \frac{\Phipw(S_2,K_2)}{\Phipw(R,K_2)} \, \left( \frac{\JS(S_2)}{\avJS} - 1 \right) \, e^{\frac{\sum_{i=1}^N \mathbf{k}_i^2}{4C}} \right\rangle & \, \\
		+ \left\langle \frac{\Phipw(S_1,K_1)\Phipw(S_2,K_2)}{\Phipw(R,K_1) \, \Phipw(R,K_2)} \, \left( \frac{\JS(S_1)}{\avJS} - 1 \right) \right.\\
	    \times \left. \left. \left( \frac{\JS(S_2)}{\avJS} - 1 \right) \, e^{2 \frac{\sum_{i=1}^N \mathbf{k}_i^2}{4C}} \right\rangle \right| & < \varepsilon \label{eq:requirement_no_sign_problem_any_orbital} \; .
	\end{split}
\end{equation}
From  Eq. \eqref{eq:requirement_no_sign_problem_any_orbital} we see that the role played by $e^{\frac{\sum_{i=1}^N \mathbf{k}_i^2}{4C}}$ is now played by
\begin{equation}
	\left\langle \frac{\Phipw(S,K)}{\Phipw(R,K)} \, e^{\frac{\sum_{i=1}^N \mathbf{k}_i^2}{4C}}  \right\rangle_{K} \label{eq:generalized_decay_factor}
\end{equation}
where $\langle \dots \rangle_{K}$ denotes the average over $K$ obtained by sampling from $\bpsifswf^*(R,S_1,K_1) \times \bpsifswf(R,S_2,K_2)$ for any given $R$, $S_1$ and $S_2$.
In other words, the factor which controls the efficiency of the calculation is now a function of $R$ and $S$.

\begin{figure}
	\centering
		\includegraphics[width=8.5cm]{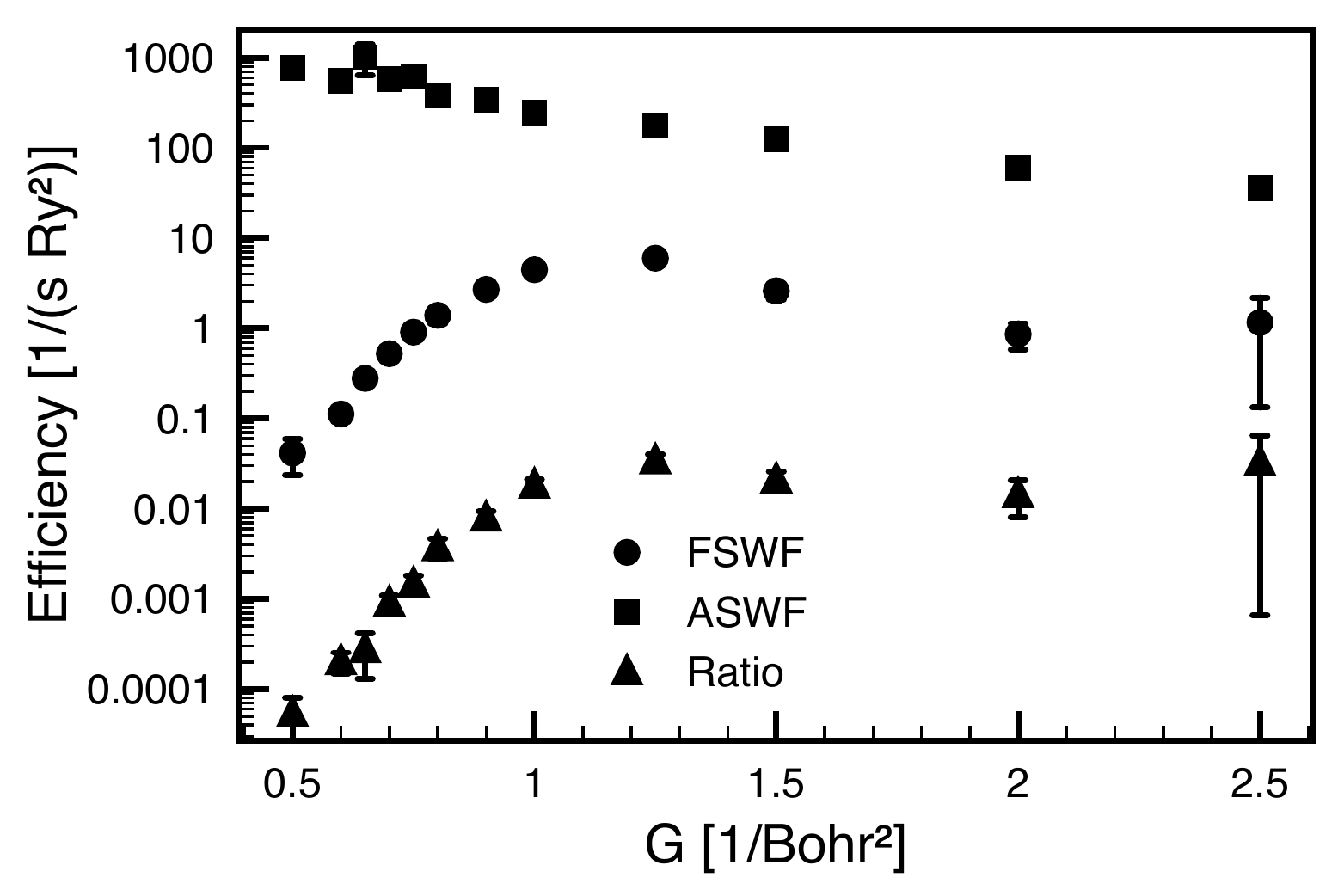} 
	\caption{Efficiency as a function of the gaussian coefficient G which enters in the Slater determinant orbitals. The dataset Ratio reports the dimensionless ratio between the FSWF and the ASWF values. Results were obtained by computing the expectation value of the energy per particle of the electronic structure of an hydrogen bcc atomic crystal with $r_s=1.7$ and periodic boundary conditions. The lattice positions required by the gaussians are the hydrogen's protons. We have used $C=1$.}
	\label{fig:Gswf_dependence}
\end{figure}

Since we have no simple interpretation of Eq. \eqref{eq:generalized_decay_factor}, we have numerically estimated its dependence on the degree of localization of the orbitals employed in the Slater determinant.
For doing so we have neglected the Jastrow terms, and considered only a kernel $\Xi$ and a Slater determinant $\Phi$ containing gaussian orbitals of the form $e^{-G(S-P)^2}$, where $P$ labels some lattice positions.
Figure \ref{fig:Gswf_dependence} shows our results.

The numerical results demonstrate that the efficiency increases exponentially when the orbitals become more localized until it reaches a maximum and finally begins to decrease.
Such decrease at large $G$ is due to other reasons than the sign problem, as it affects also the ASWF.
Therefore, in order to isolate the sign problem dependency, we have introduced the ratio between the FSWF efficiency and the ASWF one, which are represented by triangular symbols in Fig. \ref{fig:Gswf_dependence}.
Looking at these data, we can notice that at large $G$ there is a plateau rather than a decay.

The threshold at which the ratio reaches a plateau can be interpreted as the degree of localization at which the fermionic statistics become irrelevant and the quantum particles can be conveniently approximated as distinguishable.

In conclusion we have shown that there is a strong correlation between the localization of the Slater determinant orbitals and the sign problem which can be used to improve the efficiency of fermion simulations.


\section{Alleviation of the Fermion Sign Problem and Approximated Methods} 
\label{sec:alleviation}

As we have shown in the previous Sections,  Fermi statistics entail a sign problem which in general makes simulations of large number of fermions prohibitive.
Nevertheless, given the importance of fermionic systems for the comprehension of many natural phenomena (e.g. quantum chemistry), different methods which cope with these difficulties have been devised.
Here we will focus on methods which involve Monte Carlo algorithms.

A first direct approach is to try to keep $\tau$ as small as possible, and employ a very good starting trial function.
In this way, it is possible to find a lower upper bound to the exact ground state energy, systematically improvable by increasing the quality of the trial function.
This approach goes under the name of \emph{release-node} or \emph{transient estimates}
\cite{ReleaseNode,Ceperley_Alder:electron_gas, CEPERLEY07021986,PhysRevLett.46.728, prospects_release_node_QMC, Carlson_Kalos:mirror_potentials}, and it corresponds to a direct employment of the FSWF with a value of $C$ as small as possible within given computational limits.
Such methods would greatly benefit from any method able to alleviate the severity of the sign problem.
In the past years some remarkable improvements have been achieved in this direction.

\begin{figure}
	\centering
		\includegraphics[width=8.5cm]{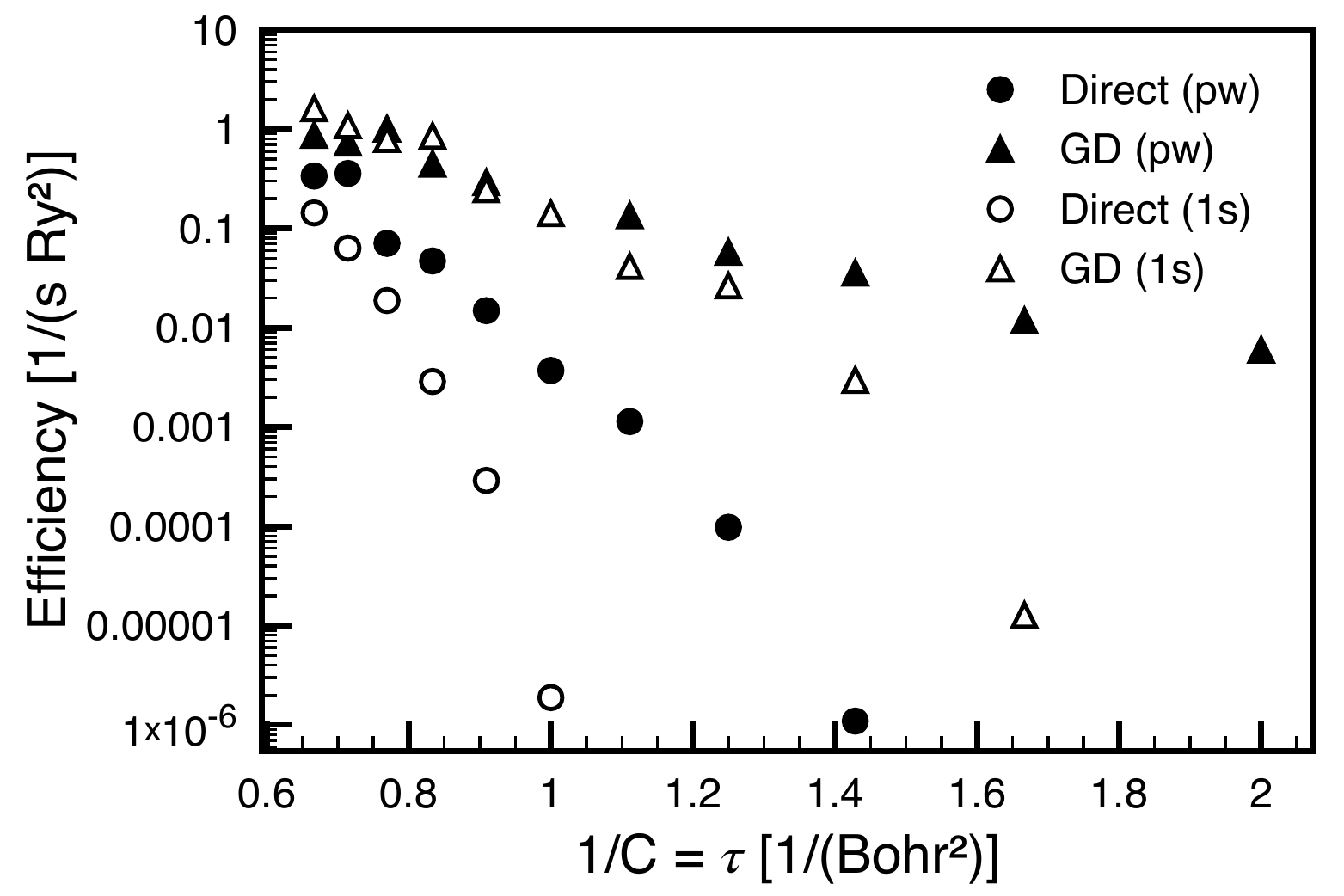} 
	\caption{Comparison between a direct simulation employing the FSWF and a calculation which employs the $\GD$ formulation. The efficiency is calculated for different $C$ and with two different choices for the orbitals embedded in the Slater determinant: The 1s orbitals and simple plane-waves (pw). To obtain these results we simulated the electronic structure of 3D solid hydrogen with a bcc crystal structure and $r_s=1.8$, employing a Yukawa Jastrow both for $\JR$ and $\JS$. The results refer to simulations done with $16$ atoms.}
	\label{fig:GD_varC}
\end{figure}

\begin{figure}
	\centering
		\includegraphics[width=8.5cm]{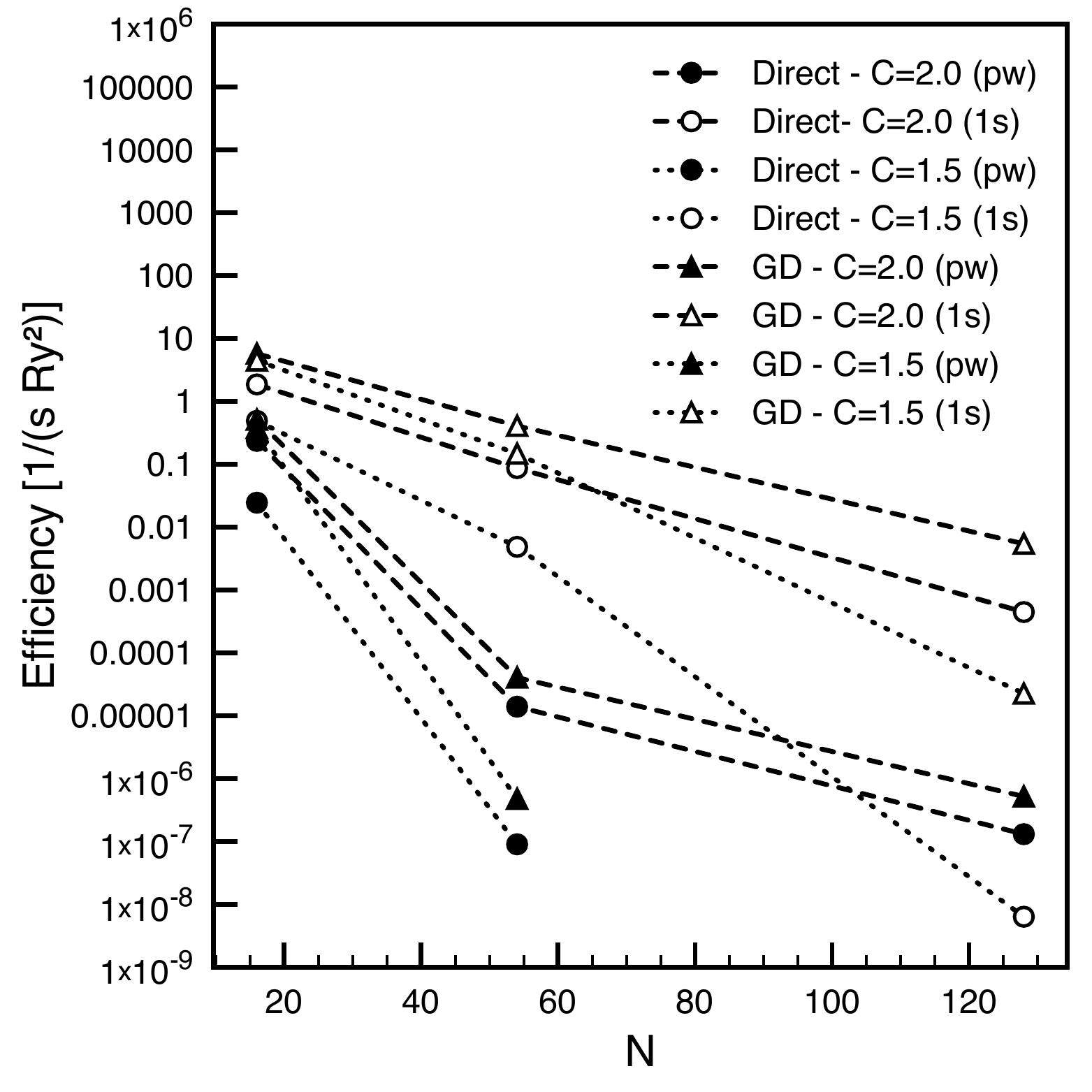} 
	\caption{Same as in Fig. \ref{fig:GD_varC} but exploring the dependency of the efficiency on $N$. We show the results for different orbitals and choices of $C$. The simulated physical system was the same as for Fig. \ref{fig:GD_varC}.}
	\label{fig:GD_varN}
\end{figure}

We begin by outlining briefly the \emph{Gaussian Determinant} ($\GD$) method \cite{PhysRevLett.102.255302}, introduced for the study of vacancies in solid ${^3\text{He}}$, and further investigated in \cite{PhysRevE.90.053304}.
The leading idea is to sum over all the possible shadow permutations by means of an anti-symmetrical kernel consisting of a determinant of gaussian orbitals.
In practice, it is sufficient to replace the kernel with
\begin{equation}
	\Xi(R,S) \; \rightarrow \; \GD(R,S) \equiv \det \left( e^{-C (\mathbf{r}_{\alpha} - \mathbf{s}_{\beta})^2} \right) \; ,
\end{equation}
where $\alpha$ and $\beta$ label rows and columns of the matrix.
Figure \ref{fig:GD_varC} and \ref{fig:GD_varN} demonstrate that the exponential pre-factor is significantly reduced.
The extension of the $\GD$ technique to PIGS is straightforward, as it is sufficient to replace each gaussian kernel with a determinant of gaussians, exactly as for the SWF.
In an unconstrained DMC method, where walkers diffuse according to the standard gaussian term, we can multiply the branching probability by
\begin{equation}
	\GD(R,R^{\prime}) / \exp{\left( - \frac{(R-R^{\prime})^2}{4 \tau} \right)} \, ,
\end{equation}
with typical values close to $1$ for small $\tau$.
A negative value of $\GD$  implies a switch of the walker from the population carrying a positive sign to the one representing negative contributions.

\begin{figure}
	\centering
		\includegraphics[width=8.5cm]{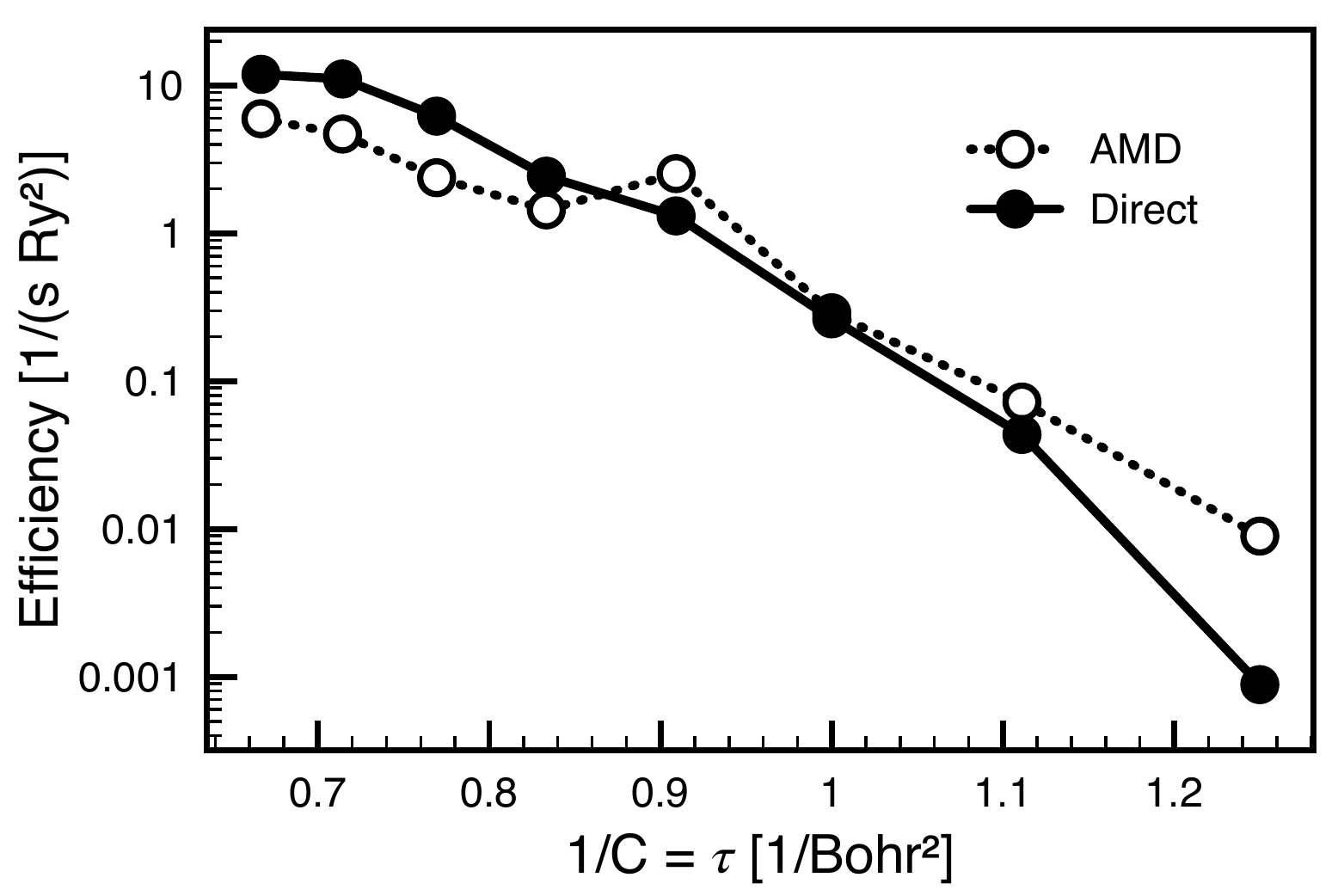} 
	\caption{Performance comparison between calculations with the FSWF using the direct algorithm and the AMD method, for different values of $C$. The results refer to the computation of the potential energy of the electronic structure of 16 hydrogen atoms in a bcc crystal structure at $r_s=1.8$. We employed a Yukawa Jastrow both for $\JR$ and $\JS$, and 1s orbitals within the Slater determinant. For approximating the marginal distribution, we used a Slater determinant embedding orbitals from Quantum Espresso \cite{QE-2009}. The AMD results refer to the best efficiency attainable by varying the parameters $\Lambda$ and $M_S$ (see \cite{PhysRevE.90.053304}).}
	\label{fig:AMD_varC}
\end{figure}

\begin{figure}
	\centering
		\includegraphics[width=8.5cm]{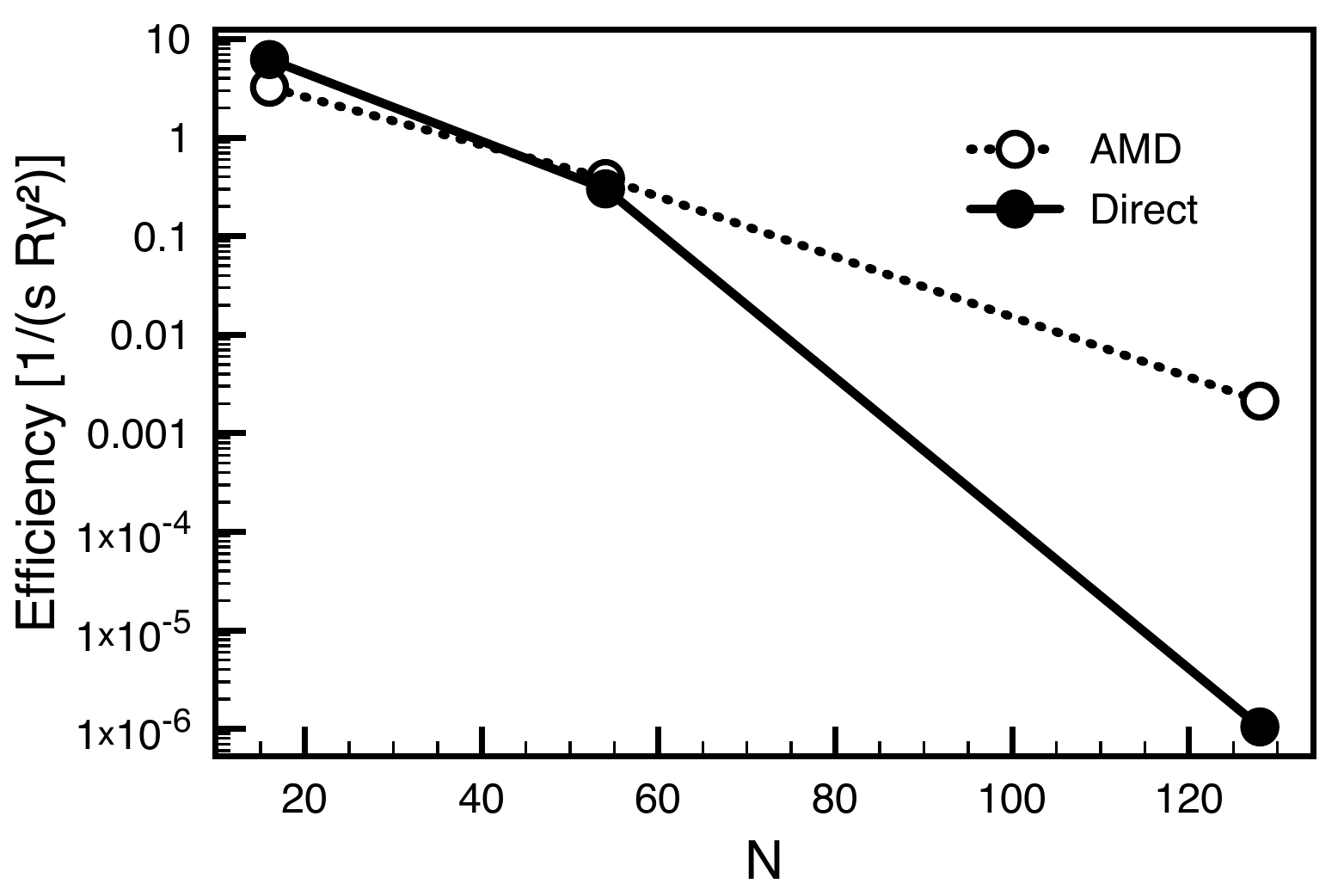} 
	\caption{Same as in Fig. \ref{fig:AMD_varC} but changing the number of simulated particles $N$. We used $C=1.3~\text{Bohr}^{-2}$.}
	\label{fig:AMD_varN}
\end{figure}

A second method is the \emph{Approximated Marginal Distribution Method} (AMD) \cite{PhysRevE.90.053304}.
Here, the leading idea is to sample the $R$ coordinates from a modified sampling distribution which is intended to account for the integration over the shadows, and therefore provides a better representation of the marginal distribution for $R$.
At the same time, the weights which carry the sign are computed by summing up the contributions coming from multiple sampling of the shadows, according to the grouping technique.
The improvements attainable with this method can be seen in Fig. \ref{fig:AMD_varC} and \ref{fig:AMD_varN}.
We point out that this method is valuable when the sign problem is "strong", whereas the direct algorithm outperforms it in the opposite limit.

Several approximated techniques which avoid the sign problem are used routinely.
Probably the most common one is the fixed-node approximation \cite{Reynolds,Anderson:fixed_nodes, kalos:fixed_nodes_boundaries, CEPERLEY07021986,Ceperley_Alder:electron_gas} which
imposes a nodal surface, forcing the solution to be antisymmetrical.
Concretely, this is accomplished by taking an antisymmetrical wave function $\psiT$ and use the DMC scheme restricted to the positive (or negative) domain of $\psiT$: Whenever a walker crosses the nodal surface, it is suppressed.
Through this procedure it is possible to filter the best antisymmetrical ground state within the given nodal surface.
Therefore, the nodal surface is the input that will determine the quality of the final result.
The fixed-node approximation has been successfully extended to Path Integral Monte Carlo \cite{PhysRevLett.69.331}.
For transferring the fixed-node method to FSWF calculations, it is sufficient to choose a nodal surface (typically using the Slater determinant of the FSWF) and to require that $S$ and its imaginary time projection $R$ are in the same nodal region.

In the following we outline three general approaches which have been devised in the past and aimed to an exact solution of the fermion sign problem.
Even if they have not been able to fully overcome the exponential behaviour, they might have the potential to alleviate this trend.

In 1982, a Green's function Monte Carlo algorithm for fermions based on a \emph{cancellation process} has been introduced \cite{arnow_kalos_lee_schmidt:fermion_green_function_MC}.
The proposed algorithm is based on the intuitive idea of having two different populations of walker, one carrying a positive sign, and the other one a negative sign, which will cancel each other if they get "close enough".
The method has been employed for few particle systems, but it suffers from a substantial drawback that prevents its application to many-body system.
The reason behind this is that the algorithm requires a high density of walkers, and unfortunately such requirement implies an exponential growth of the computational cost proportional to the number of simulated particles, because of the increased dimensionality of the problem.
Further works in the same direction \cite{PhysRevLett.67.3074, PhysRevE.50.3220, Pederiva1999440, PhysRevLett.85.3547} have
shown that calculations for small system sizes are feasible, but the fermion sign problem is not solved in general.

In 1985 a method for treating the fermion sign problem involving \emph{mirror potentials} has been devised \cite{Carlson_Kalos:mirror_potentials}, where a fictitious repulsive interaction is used to keep the distribution of positive and negative walkers apart from each other, and hence avoids the collapse into the same bosonic ground state.
By increasing the repulsion between the two populations this method reduces to the fixed-node approximation, whereas when it is set to zero one obtains a release-node simulation.
The mirror potential method allows exact fermion calculations, but is limited to a small number of particles due to the exponential growth in number of walkers required to describe the mirror potential, similarly to what happened with the cancellation idea.
To overcome this  difficulty, it is possible to make use of a trial wave function.
However this will lead to an approximate result, although potentially more accurate than the fixed-node one.

In the context of Path Integral Monte Carlo, there has been an interesting attempt towards the solution of the fermion sign problem in 1998-2000 \cite{PhysRevLett.81.4533, PhysRevE.61.5961}.
The proposed approach goes under the name of \emph{multilevel blocking}, and  consists of distributing the integrals for the propagation of a single imaginary time step $\Delta \tau$ in an elaborated pyramidal structure, solving it in a bottom-up fashion.
First one computes the $\Ntau/2$ integrals at the bottom, and then use these informations to compute the integrals at a coarser level, i.e. for $2 \Delta \tau$.
This procedures is repeated until the integral for the full imaginary-time propagation $\Ntau \Delta \tau$ is found.
However, we have seen that even a single integration step already introduces a sign problem which scales exponentially in the number of particles, so that the proposed scheme will eventually scale exponentially, too.

In 2009 a new method, FCIQMC \cite{ali_alavi_fermion_MC}, for treating fermion very accurately has been introduced, based on a Monte Carlo imaginary-time projection technique performed in the space of Slater determinants, in the spirit of Full Configuration Interaction (FCI).
In this method, if the number of walkers is sufficient to populate such space, the FCI wave function will emerge from the calculation within a less severe computational cost compared to the traditional one\cite{Bryan}.
However, the use of such variational space for treating particle correlation implies a size-extensivity problem.
Nevertheless, this method has demonstrated to be competitive with other highly-accurate methods employed in quantum chemistry \cite{a_alavi:JCP2011, PhysRevB.86.035111, a_alavi:JCP2010,PhysRevB.85.081103, a_alavi:solids}.

\section{Computational Complexity of the Fermion Sign Problem}
\label{sec:complexity}
In this section we discuss the computational complexity of the fermion sign problem in QMC.
Before going into the details, we would like to briefly introduce the reader to the complexity classes P, NP, NPC, and NP-hard.
The interested reader can refer to \cite{aho1974design,garey1979computers,book:algorithms} for an exhaustive introduction to the topic.

P is the class of problems which are solvable in polynomial time: Provided an input problem of size $n$, it exists an algorithm which can solve it in $O(n^k)$ time where $k$ is a constant.
In our specific case, the input is provided by the physical parameters of the system, the variational parameters to be employed, and the Hamiltonian of the system.
The problem's size is given by the number of simulated particles.

NP (Nondeterministic Polynomial) problems are the ones which can be verified in polynomial time: Given a solution (\emph{certificate}) of the problem, it exists an algorithm that can verify its correctness in $O(n^k)$ time.
We remark that any problem in P belongs also to NP, because if it is possible to solve a problem in polynomial time, such a solution can be used to verify a certificate.

In order to illustrate the NPC and the NP-hard classes we need to further introduce the concept of \emph{reducibility}.
The problem A is said to be \emph{polynomial-time reducible} to problem B if it is possible to find a map $A \mapsto B$ which is computable in polynomial time.

The NPC (\emph{NP-Complete}) problems are the NP problems which have the additional property of being polynomial-time reducible to any other NP problem.
If a problem possesses the latter property but not necessarily the first one, then it is said to be \emph{NP-hard}.


In the work of Troyer and Wiese \cite{PhysRevLett.94.170201}, it has been shown that a general QMC algorithm which is able to compute the most general partition function can also  be used to solve a NPC problem. The reference NPC problem \cite{0305-4470-15-10-028} is the following: Given a classical 3D Ising spin glass with Hamiltonian
\begin{equation}
   H = - \sum_{<i_1,i_2>} J_{i_1 i_2} S_{i_1} S_{i_2} \label{eq:hamiltonian_spin_glass}
\end{equation}
and a bound energy $E_0$, does a spin configuration with energy $ E \leq E_0$ exist?
The interaction matrix $J$ has values $j$, $0$, or $-j$ chosen randomly, and the spins $S$ can have values $\pm 1$.

The latter decision problem is connected to Monte Carlo calculations by the fact that provided a large enough inverse temperature $\beta$, the average energy of the spin glass system will be less than $E_0 + j/2$ if a configuration with $E \leq E_0$ exists, and larger than $E_0 + j$ otherwise (basically, the simulated annealing minimization method).
As a consequence, the computation of E with a Monte Carlo simulation would provide an answer to the given 3D Ising spin glass NPC problem.
In other words, such Monte Carlo average is necessarily at least as hard as a NPC problem, i.e. it is NP-hard.

Since the classical NPC problem in Eq. \eqref{eq:hamiltonian_spin_glass} can be mapped in a quantum one simply by replacing classical spins with quantum ones, a general algorithm to solve quantum problems (including those with a sign problem) will provide also a solution to our NPC problem and thus will be NP-hard \cite{PhysRevLett.94.170201}.
However, not all many-fermion systems pose NP hard problems and it remains open if there is a criterion that allows us to immediately distinguish a NP-hard situation from P or BPP.

It would be of great importance to be able to better characterise the fermion sign problem complexity.
Is it possible to devise a criterion in order to predict when the QMC simulation will be NP-hard?
Is it possible to find a case in which the NP-hardness originate from the fermion statistic?
Some progress in this direction has been made recently \cite{PhysRevB.48.589,PhysRevB.71.155115,PhysRevB.89.111101,PhysRevLett.115.250601,2016arXiv160105780L,2016arXiv160101994W} identifying sets of Hamiltonians wihout sign problem.

In the following we discuss a definition of the fermion sign problem concerning
purely continuum QMC simulation with local interactions, using the SWF formalism.

\emph{Decisional fermion sign problem}: Is the value of the integral
\begin{equation}
\int dS \, e^{-C (R-S)^2} \det \left( \phi_{\alpha}(\mathbf{s}_{\beta}) \right) \, \JS(S)  \label{eq:fermion_sign_problem}
\end{equation}
strictly greater than zero?

If one could answer to this decisional problem, exact fermion calculations in polynomial time would be possible  
by employing the fixed-node algorithm.
If the provided answer is affected by a statistical error, as it would be the case by using a Monte Carlo technique, then the fixed-node method will have to make use of the penalty method \cite{ceperley:penalty_method}, whose efficiency will decrease exponentially with the given statistical error. 

Notice that this decisional problem does not admit the existence of a certificate, as it is not in the form "does ... such that ... exists?".
Therefore one should find a corresponding problem which will allow to identify it as a NP-hard problem (if it is such).
Unfortunately, we have to leave this question open.

\section{Conclusion} 
\label{sec:conclusion}

We have presented the SWF formalism and showed that the fermion sign problem appearing in typical imaginary-time projection continuum QMC methods can be reduced to the sign problem of the Fermionic Shadow Wave Function.
This formalism was used to characterise both analytically and numerically the fermion sign problem, demonstrating its dependence on the number of particles, length of the imaginary-time projection, and localisation of the system.
Even though it seems that the exponential decay of the simulation efficiency cannot be overcome, we have shown that some methods can lead to a significative reduction of its exponential factor, thus extending the applicability of exact QMC methods for fermions.
Concerning the complexity class of imaginary-time projection QMC algorithms, a separate proof of the NP-hardness of 2D and 3D fermionic systems with local interactions in continuum space is still lacking.

\begin{acknowledgments}
F.C. would like to acknowledge the Nanosciences Foundation of Grenoble for financial support and T. D. K\"uhne for allowing us to access  the Mogon HPC which has been used for most of the numerical calculations.

\end{acknowledgments}

\bibliographystyle{apsrev}
\bibliography{origin_sign_problem}

\end{document}